\documentclass[pop, fleqn, 12pt]{w-art}
\usepackage{times}
\usepackage{w-thm}
\usepackage{graphics,amsmath,amssymb,amscd,cite}
\usepackage{graphicx}

\newtheorem{Theorem}{Theorem}[section]
\newtheorem{Lemma}[Theorem]{Lemma}
\newtheorem{Proposition}[Theorem]{Proposition}
\theoremstyle{definition}
\newtheorem{Definition}[Theorem]{Definition}
\theoremstyle{remark}

\newcommand{\A}{{\mathbb{A}}}
\newcommand{\Spec}{{\rm Spec}}

\newcommand{\bp}{\begin{Proposition}}
\newcommand{\ep}{\end{Proposition}}
\newcommand{\bl}{\begin{Lemma}}
\newcommand{\el}{\end{Lemma}}
\newcommand{\bt}{\begin{Theorem}}
\newcommand{\et}{\end{Theorem}}
\newcommand{\bd}{\begin{Definition}}
\newcommand{\ed}{\end{Definition}}

\newcommand{\Aut}{\rm{Aut}}

\DeclareFontFamily{U}{rsf}{}
\DeclareFontShape{U}{rsf}{m}{n}{<5> <6> rsfs5 <7> <8> <9> rsfs7 <10-> rsfs10}{}
\DeclareMathAlphabet\Scr{U}{rsf}{m}{n}

\def\Z{{\mathbb Z}}
\def\C{{\mathbb C}}

\newcommand{\be}{\begin{equation}}
\newcommand{\ee}{\end{equation}}
\newcommand{\bea}{\end{eqnarray}}
\newcommand{\eea}{\end{eqnarray}}

\begin{document}
\title[Non-commutative moduli spaces of topological D-branes]
{Non-commutative moduli spaces of topological D-branes}

\author[Calin Iuliu Lazaroiu]{Calin Iuliu Lazaroiu\inst{1,}%
  \footnote{\quad E-mail:~\textsf{calin@maths.tcd.ie}}}
\address[\inst{1}]{Department of Mathematics\\Trinity College Dublin\\Dublin 2, 
Ireland\\calin@maths.tcd.ie}

\maketitle

\begin{abstract}
We give a general construction of extended moduli spaces of
  topological D-branes as non-commutative algebraic varieties. This shows that
noncommutative symplectic geometry in the sense of Kontsevich arises naturally
in String Theory.
\end{abstract}

\vskip .6in


\section{Introduction} 

The moduli space of topological D-branes provides a first
approximation to the quantum geometry seen by open strings. Because
the algebra of boundary operators on a topological open string
worldsheet is noncommutative, one tests quantum geometry with a
noncommutative probe, so one expects a non-commutative description of
the relevant moduli space.

In the traditional approach to D-brane geometry, one starts by fixing
a Chan-Paton representation $\Gamma$ and constructs a commutative
moduli space which depends markedly on $\Gamma$. This dependence makes
it difficult to identify the deeper structure which lies behind
particular examples. The task of extracting the latter requires a
universal formulation of the deformation problem for topological
D-brane systems, which describes boundary deformations in a
representation-agnostic manner.  In \cite{nc}, we showed how one can
carry out this task starting from the most basic conditions on
topological string amplitudes, known as the $A_\infty$ constraints. In
fact, we carried out this analysis for a general system containing a
finite number of D-branes. Using an algebraic formulation of such
systems, we showed that topological boundary amplitudes can be packed
into a `noncommutative generating function' $W$ which generalizes the
generating function considered in \cite{HLL} (itself a generalization
of the D-brane superpotential of \cite{CIL4}). Formally, $W$ is an
element of the zero'th (relative) Karoubi complex of a certain
associative superalgebra $A$, and can also be viewed as a linear
combination of necklaces on a superquiver. Moreover, the boundary
topological metrics of the system induce a noncommutative symplectic
form on $A$ in the sense of \cite{Konts_formal}, which we denote by
$\omega$. It was shown in \cite{nc} that $W$ is subject to certain
constraints, which naturally encode a series of complicated algebraic
conditions on the tree-level scattering amplitudes between the open
topological strings present in the system. Using these constraints,
one can recover $\omega$ from $W$, which means that $W$ suffices to
completely characterize tree-level dynamics.  The construction of the
extended moduli space proceeds in algebraic manner, by considering the
invariant part $(A/J)^{\cal G}$ of the quotient of $A$ through the
'noncommutative critical ideal' $J$ of $W$ under the action of a
certain group ${\cal G}$ which encodes all autoequivalences of the
collection of boundary amplitudes. This algebraic construction has a
non-commutative-geometric interpretation in which $A$ is viewed as the
noncommutative coordinate ring of a noncommutative affine space $\A$,
allowing one to view $W$ as a noncommutative function (in the sense of
\cite{Konts_formal}) defined on that space. Then $A/J$ is the
noncommutative coordinate ring of the `critical locus' ${\cal Z}$ of
$W$, viewed as a subspace of $\A$, while $(A/J)^{\cal G}$ is the
coordinate ring of the (noncommutative) extended moduli space of
interest, which we will denote by ${\cal M}$.

\section{The $A_\infty$ structure of finite topological D-brane systems}

It is well-known that scattering amplitudes of open strings
can be encoded by a series of multilinear maps (the so called {\em boundary string products}) 
defined on the space $E$ 
of boundary operators and taking values $E$. As explained in detail
in \cite{HLL} (see also \cite{ Costello1, Costello2, Stasheff_Kajiura}), 
the hierarchy of contact terms arising when two boundary operators
approach each other is constrained by the geometry of the natural
compactification of the moduli space of disks with boundary punctures. As a
consequence, the boundary scattering products of a topological open string
theory are subject to a countable series of constraints, which turn out to
coincide with the algebraic conditions defining an $A_\infty$ algebra
\cite{Stasheff}, a fact anticipated in \cite{Gaberdiel} in the context of
open bosonic string field theory. 
Generalizing this observation to {\em systems} of topological D-branes leads to the
following conclusion:

{\em The collection of all (integrated) boundary and boundary-condition
  changing tree-level amplitudes of a topological string theory forms a weak,
  cyclic and unital $A_\infty$ category.}

Above, the worldsheet model of a topological string theory 
is assumed to be a (possibly massive) deformation of 
a topological worldsheet theory obtained by twisting a model
with N=2 superconformal invariance. Indeed, 
the proof of \cite{HLL} relies on the existence of a point in the topological deformation
space at which the worldsheet model is a topological conformal field theory. 
We refer the reader to \cite{HLL} for a precise formulation of the
assumptions. A standard example are the topological open string theories 
associated with topological sigma models \cite{Witten_mirror, Witten_CS}.

\section{Encoding the $A_\infty$ structure}

As usually formulated, the structure of a weak, cyclic and unital $A_\infty$
category is rather complicated (see Appendix A to \cite{nc}) 
and does not easily allow for general insights. When the system contains a
finite number of topological D-branes,  of the results of 
\cite{nc} is an equivalent description of this structure in a language which
clarifies its non-commutative geometric meaning. For this, we start with the 
observation that a finite collection ${\cal Q}_0$ of D-branes -- viewed as 
an abstract finite set-- allows us to define a semisimple commutative algebra $R$
via the following expression: 
\be
R:=\oplus_{u\in {\cal Q}_0}{\C \epsilon_u}
\ee
where $\epsilon_u$ are mutually commuting idempotents: 
\be
\epsilon_u\epsilon_v=\delta_{uv} \epsilon_u~~.
\ee
Noticing that $R$ is unital with unit $1_R=\sum_{u\in {\cal Q}_0} \epsilon_u$, we find
that the field of complex numbers embeds into $R$ via: 
\be
\C \equiv \C 1_R \subset R~~.
\ee
For each ordered pair of D-branes $(u, v)$, we have a super-vector space $E_{uv}$ spanned
by all topological boundary operators which change the boundary condition from $u$
to $v$ (the case $u=v$ corresponds to the space of boundary operators for a topological
open string whose endpoints lie on the D-brane $u$). The {\em total boundary
  space} is defined as the super-vector space
$E=\oplus_{u,v\in {\cal Q}_0}{E_{uv}}$. A useful observation is that giving such a
decomposition amounts to giving an $R$-superbimodule structure on
$E$. Thus the `boundary sector decomposition' of the system is encoded by an 
$R$-superbimodule $E$ (which in most physical examples is finite-dimensional
as a complex vector space). 

The next step is to consider the tensor algebra $A=T_R E[1]^{\rm v}$ of the 
parity changed dual of this super-bimodule. This is an $R$-superalgebra since
it contains $R$ as a subalgebra in its even subspace. Applying a super-extension of a basic construction 
to this superalgebra allows us to construct its (relative) differential envelope $\Omega
_R A$, a $\Z\times \Z_2$-graded associative algebra whose elements are known 
as noncommutative differential (super)forms with coefficients in $A$. The envelope  
$\Omega_R A$ is endowed with a differential $d$, which in the conventions of 
\cite{nc} has bi-degree $(1,0)$. The {\em relative Karoubi complex} of $A$ is the 
quotient $C_R(A):=\Omega_R A/[\Omega_R A, \Omega_R A]$ of $A$ via the sub{\em
  space} $[\Omega_R A, \Omega_R A]$ defined as the image of the graded
commutator map $[.,.]:\Omega_R A\times \Omega_R A\rightarrow \Omega_R A$. This
is a $\Z\times \Z_2$-graded complex, whose differential ${\bar d}$ is
induced from $\Omega_R A$. The elements of the $\Z$-homogeneous components $C^n_R(A)$ are
called non-commutative differential $n$-superforms with  coefficients in
$A$. According to \cite{Konts_formal}, a {\em non-commutative symplectic form} on $A$
is an element $\omega\in C^2_R(A)$ which satisfies a certain non-degeneration
condition (see \cite{nc} for a precise formulation of this
property). This noncommutative symplectic form has a $\Z_2$-degree ${\tilde \omega}$,
which is one of the data of the underlying topological string theory.
As explained in \cite{nc},  $\omega$ induces a Lie superbracket on the
parity-changed space $C^0_R(A)[{\tilde \omega}]$. We are ready to formulate\footnote{This is 
also derived in \cite{Lazarev} in the particular case when the left and right
$R$-module structures on $E$ agree.} one of the results of
\cite{nc}. 

{\em Giving a cyclic $A_\infty$ category with  finite-dimensional morphism spaces 
and having a finite number of objects is equivalent to giving the following
data: 

(1) A finitely-generated semisimple commutative algebra $R$ over the complex
    numbers.

(2) An $R$-superbimodule $E$ which has finite dimension as a vector space over 
$\C\subset R$

(3) A noncommutative symplectic form $\omega\in C^2_R(A)$ on the
tensor algebra $A:=T_R E$, whose $\Z_2$-degree we denote by ${\tilde \omega}$.

(4) An element $W\in C^0_R(A)$, of $\Z_2$-degree ${\tilde \omega}+1$, which
satisfies the condition $\{W,W\}=0$, where $\{.,.\}$ is the Kontsevich superbracket defined by $\omega$. }

Using the non-commutative geometric interpretation of \cite{Konts_formal},
this characterization relates finite D-brane systems to a non-commutative
version of the QP-manifolds of \cite{Konts_Schwarz}

Unitality of the weak $A_\infty$ category of boundary string products can also
be encoded in this language. For this purpose, one notices that any vector
$x\in E$ defines a linear operator $\delta_x:C^0_R(A)\rightarrow A$, which we shall
call  the (left) {\em cyclic derivative}  along $x$ (indeed, it is shown in \cite{nc} that this is
a generalization of the cyclic derivatives of \cite{RSS, Voiculescu}). 
Then a result of \cite{nc} characterizes unitality of the weak $A_\infty$ category of
boundary string products as the requirement that there exists an 
even central element $\lambda$ of $E$ (called the {\em total $A_\infty$ unit}) 
such that $\delta_\lambda W$ belongs to a
certain subspace  of $A$, the so-called subspace of noncommutative
moment maps. One can also  show that a noncommutative 
generating function satisfying this property can be used to reconstruct the 
noncommutative symplectic form $\omega$.

\section{The extended noncommutative moduli space} 

The (extended) noncommutative moduli space of the finite topological D-brane 
system can be cosntructed follows. First, one defines the {\em noncommutative
  critical ideal} $J$ of $W$ to be the two-sided ideal generated by all 
cyclic derivatives $\delta_x W$, where $x$ runs over $E$. Since 
$\delta_x$ depends linearly on $x$ (see \cite{nc}), this ideal is finitely
generated:any basis $e_i$ of $E$ gives a system of generators 
$\delta_{e_i} W$. Further, one considers the group ${\cal G}$ formed by all
unital autoequivalences of the underlying weak $A_\infty$ category which preserve the
cyclic structure given by the boundary topological metrics. As shown in
\cite{nc}, ${\cal G}$ can also be described as the group formed by those
{\em super}algebra automorphisms of  $A$ which preserve the noncommutative symplectic form 
$\omega$ and obey a certain splitting
property with respect to the $A_\infty$ unit $\lambda$. Using these objects,
we consider the invariant subalgebra $(A/J)^{\cal G}$ of the quotient 
$A/J$. 

To interpret this geometrically, we follow \cite{Konts_formal} in viewing 
$A$ as the noncommutative coordinate ring of a noncommutative symplectic space $\A$. 
With this interpretation, the quotient $A/J$ is the noncommutative
coordinate ring of a noncommutative subspace ${\cal Z}$ of $\A$, and the group ${\cal
  G}$ acts on $\A$ via symplectomorphisms. The algebra  of invariants 
$(A/J)^{\cal G}$ is taken to describe the quotient of ${\cal Z}$ through
this action of ${\cal G}$, which defines the noncommutative extended moduli 
space ${\cal M}$. With this interpretation, $(A/J)^{\cal G}$ is the
noncommutative coordinate ring of the desired extended noncommutative moduli space.

\section{Representations} 

It can be shown that Chan-Paton representations of the underlying open strings
(which are specified by picking multiplicities for each $D$-brane in the system)
amount to representations of the tensor superalgebra $A$. Using this relation,
one finds that the  extended version of the commutative moduli space
considered in traditional analyses of the moduli space of D-brane systems
arise as the collection `matrix-valued' points of ${\cal M}$. Therefore,
the noncommutative moduli space ${\cal M}$ plays the role of a universal 
moduli space which encodes all choices of D-brane multiplicity. This gives a 
physical realization of the ideas of Kontsevich \cite{Konts_formal}, which 
were extensively developed in \cite{Ginzburg, BEV, VG,
  Ginzburg_lectures,LB_quivers, Bergh, LB_lectures}.

\section{On the computation of ${\cal M}$} 

It is in general straightforward to describe ${\cal Z}$ explicitly upon using
standard algebraic methods. The computation of ${\cal M}$ is typically much
more difficult, due to the fact that it might be quite nontrivial to explicitly
determine the group ${\cal G}$. Indeed, ${\cal G}$ arises as a certain subgroup
of $\Aut_R^\omega(A)$, the group of (relative) symplectomorphisms of the noncommutative
symplectic space $\A$ (which can be viewed as a noncommutative
space over $\Spec(R)$). The later is itself a subgroup of the group $\Aut_R(A)$ of relative 
superalgebra automorphisms, which is nontrivial to determine in general form (see
\cite{Umirbaev, DY}) even in the basic case of a single D-brane sector, for
which $A$ is a free superalgebra.  Surprisingly, this gives a direct
connection between the physics of topological D-branes and certain basic problems
in noncommutative algebra, such as the characterization of wild automorphisms. 
In spite of such difficulties, the space ${\cal M}$
has  been determined  explicitly at least in certain low-dimensional examples
\cite{nc}. 

\begin{acknowledgement}
The author is supported by the European Commision FP6
program MRTN-CT-2004-005104, in which the author is
associated with Trinity College Dublin.
\end{acknowledgement}


\end{document}